# Bound states of tunneling electrons in molecular wires

M.A. Kozhushner[1], V.S. Posvyanskii[1] and I. I. Oleynik[2]

[1]*Institute of Chemical Physics, Russian Academy of Sciences, Kosygin St., 4, Moscow 119991*
[2]*Department of Physics, University of South Florida, Tampa, FL 33620*

We have studied the bound states of the extra electron in a molecular wire, corresponding to the physical situation of the electron tunneling through the wire. Based on formalism of the scattering operator that accounts for many-electron interactions we have shown that the energy spectrum of the bound states of the extra electron in a linear chain of three-dimensional scattering centers substantially deviates for spectrum of independent single-electron states of Bloch electrons in a system described by a single-electron effective potential. In particular, in the case of elemental wire consisting of the centers with one electronic state per center, the number of states in the band can be less than number of centers of the wire. These states form two bands instead of single band and the bands exhibit non-analytical, square root dispersion $\varepsilon - \varepsilon_m \propto \sqrt{k - k_m}$ in the vicinity of the point $k_m$ where both bands come together. These findings are important for developing physically viable theory of electron tunneling in single molecular systems because the structure of spectrum of electron's bound states and its relative position in respect to the Fermi levels of electrodes determine specific mechanisms of electron transport through a molecular wire.

In recent years one-dimensional molecular structures have been actively studied both theoretically and experimentally due to their potential applications in nanoelectronics[1-3]. Various device concepts are continuing to be explored including rectifying diodes, field effect transistors and switches built from such materials as semiconductor and metal nanowires, nanotubes, small organic molecules and large biomolecules[2]. In order to fully utilize the unique properties of these one-dimensional nanostructures, a fundamental understanding of conduction mechanisms has to be achieved.

Recent experiments investigated conductivity in organic molecules, including DNA strands, and showed threshold or non-threshold behavior of conductance as a function of applied voltage[4-7]. Surprisingly, the absolute values of the conductance were found to be very high, approaching those observed in metallic wires. These experimental facts can be consistently rationalized by introducing a conduction mechanism based on resonant tunneling through the energy levels of the extra electron in the molecular bridge[8-9] or in chemical language, the energy levels of a negative molecular ion[8]. Obviously, the structure of the energy spectrum is of fundamental importance for elucidation of electron transport mechanisms in molecular wires.

Usually, the electron orbitals of a neutral molecule are calculated and used for interpretation of the transport mechanisms. This approach works for the case of extended systems with large number of electrons such as solids, where the electron spectrum is hardly perturbed when an extra electron is added to the system. However, it is well known in case of systems with small number of electrons that the energy spectrum is dramatically altered when an extra electron is introduced to the system. In case of molecules, for example, electron affinity does not coincide with the energy of the electron state next to HOMO of neutral molecule.

In experiments on electron conductance through a single organic molecules the electron tunneling is the major mechanism of electron transfer. This tunneling electron is an extra electron that interacts with the neutral molecule including its electrons and nuclei in the course of electron transition. Therefore, its energy spectrum at negative energies (in respect to vacuum level) corresponds to the energy levels of the "negative ion", i.e. bound energy spectrum of one electron plus a neutral molecular wire.

The many-body effects in the course of electron transfer through the molecule play an important role and may result in substantially different physical characteristics of electron spectrum. Therefore, the main purpose of this paper is to describe the fundamental properties of bound state energy spectrum of an extra electron tunneling through one dimensional molecular and show that due to many-body effects the properties of the bound states are drastically different from well-known properties of Bloch states of an electron that interacts with the molecule via effective single particle electron potential.

According to the Bloch theorem[10], the electronic states of the system with a periodic potential are described as the Bloch waves: the plane wave $\exp(i\mathbf{kr})$ multiplied by a function $u_\mathbf{k}$ with the period of the crystalline lattice[9]. For simplicity, let us consider a simple model of one-dimensional molecular wire, the linear chain of individual centers, see Fig. 1a. The centers might represent the functional groups or structural building blocks of the molecule. The energies of the states constituting a band of one-dimensional linear wire are the function of crystalline momentum $k$ which spans the first Brillouin zone: $-\pi/d \le k \le \pi/d$ in case of the system with periodic boundary conditions or $0 \le k_s = \pi/d \cdot n/N \le \pi/d$, $n = 1,...,N$ in the case of the wire with free-standing ends. The crystalline momentum is the quantum number to account for translational invariance of the system and the number of quasi-discrete values of $k$ is equal to $N$, the number of centers in the system.

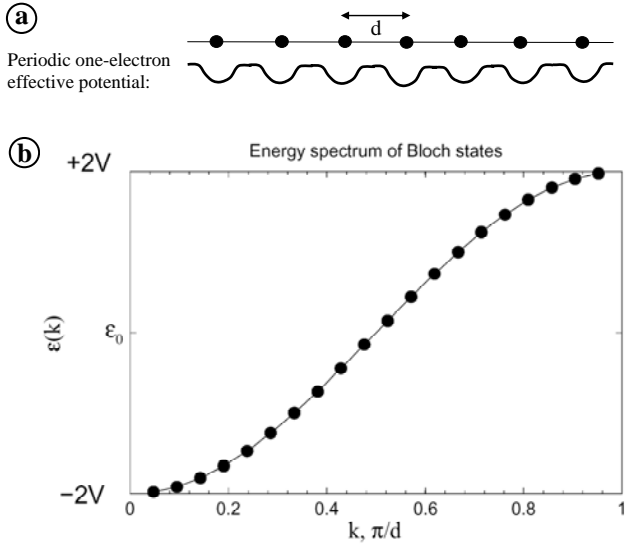

Fig. 1. a: One-dimensional periodic chain of atoms plus one-electron effective potential model; b: tight-binding energy spectrum: the number of states in the band is equal to $N = 20$, the number of centers in the wire.

In the case of elemental wire, i.e. the one-dimensional system of $N$ equally spaced potential centers with one electronic state per center, the Bloch theorem and the single electron effective potential approximation allows the wave function of an extra electron to be expressed in tight-binding representation as a linear combination of the electron wave functions of the individual centers $\varphi(\mathbf{r} - \mathbf{R}_l)$

$$\psi_k = \frac{1}{\sqrt{N}} \sum_l \sin(kR_l) \varphi(\mathbf{r} - \mathbf{R}_l) \qquad (1)$$

where $l$ is the index of each center in the wire and $\mathbf{R}_l$ is its coordinate. The interaction of the extra electron with the individual centers produces splitting of $N$ identical energy levels $\varepsilon_0$ into a band of $N$ collectivized states:

$$\varepsilon(k) = \varepsilon_0 - 2V \cos(kd) \qquad (2)$$

where wave number is $k = \pi s/(N+1)d$, $s = 1,...,N$ and $V$ is the hopping integral of the extra electron between the centers separated by distance $d$, see Fig. 1b. According to this standard picture, the number of states in the band is equal to number of centers in the wire, see Fig. 1b.

The basic assumption used to derive the tight-binding dispersion law (2) is the notion of a *single electron* interacting with the molecular wire via one-electron effective potential which is periodic with period $d$ [10], see Fig. 1a. In reality, this extra electron that participates in electron transport through the wire strongly interacts both with nuclei and other electrons of the system. In the region of negative energies below the vacuum level, i.e. in the region of energies corresponding to the tunneling regime, the many-electron effects due to the exchange interaction become very important. Therefore, the basic assumption of single-electron effective potential approximation is not strictly satisfied and it is reasonable to expect deviations from the well-established picture of electron energy spectrum.

In order to address this problem, we have developed a theory of sub-barrier electron scattering that specifically takes into account the many-electron effects including exchange interaction[9,11]. The method uses an equivalent formulation of quantum mechanics based on the formalism of scattering operators which coarse-grains the many-body effects within a one-electron framework thus resulting in an efficient solution of the tunneling problem for complex molecular systems.

We consider the case of an extra electron that tunnels through the molecule. It interacts with the nuclei and electrons of the wire in the course of the tunneling transition. We start with the notion of sub-barrier scattering that is the scattering of the electron by individual centers comprising the molecule at negative energies below the vacuum level (the zero energy is chosen to be at vacuum level). The interaction of the tunneling electron with an individual scattering center is described by the sub-barrier scattering operator $t(\varepsilon, \theta)$ which can be written in general form as[12]

$$t(\varepsilon, \theta) = \frac{2\pi b(\theta)}{\varepsilon - \varepsilon_0} + t_{pot}(\varepsilon, \theta) \qquad (3)$$

where $\varepsilon_0$ is the energy of the bound state of the isolated center, $\varepsilon$ and $\theta$ are the energy and scattering angle of the electron. The first and second terms in (3) represent the pole and potential terms of the scattering operator[12].

The sub-barrier scattering operator $t(\varepsilon, \theta)$ is calculated by using a variational asymptotic method that allows the exponential tail of the wave function of the system of $n_c + 1$ electrons ($n_c$ electrons of the scattering center plus one tunneling electron) to be determined by varying the total energy functional. During this variational procedure, the many electron wave function of the system of $n_c$ electrons is written as a Slater determinant, thus allowing the many-electron effects to be included explicitly in the calculations. The scattering operator $t(\varepsilon, \theta)$ at $\varepsilon < 0$ is obtained as the analytic continuation of $-2\pi a(\varepsilon, \theta)$ at $\varepsilon > 0$ to the region of imaginary monenta $k = i\kappa = i\sqrt{2|\varepsilon|}$, where $a(\varepsilon, \theta)$ is the standard scattering amplitude obtained from asymptotics of an electron wave function. Then, the total Green's function of the tunneling electron can be written as[11]

$$G(\mathbf{r},\mathbf{r}';\varepsilon) = G_0(|\mathbf{r}-\mathbf{r}'|;\varepsilon) + G_0(|\mathbf{r}-\mathbf{R}|;\varepsilon) t(\varepsilon,\theta) G_0(|\mathbf{R}-\mathbf{r}'|;\varepsilon) \qquad (4)$$

where **R** is the coordinate of the scattering center. The vacuum Green's function at negative electron energies $\varepsilon$ (i.e. below the vacuum level) is

$$G_0(|\mathbf{r}-\mathbf{r}'|;\varepsilon) = -\frac{1}{2\pi|\mathbf{r}-\mathbf{r}'|}\exp(-\kappa|\mathbf{r}-\mathbf{r}'|) \quad (5)$$

where $\kappa = \sqrt{2|\varepsilon|}$. We use the atomic units $e = \hbar = m_e = 1$. The expression (4) for the tunneling Green's function has clear physical meaning: the electron tunnels as free plus scattered off the center located at **R**.

The sign of $t(\varepsilon,\theta)$ is the indicator of the character of the effective interaction between the extra electron and the scattering center: $t(\varepsilon,\theta) < 0$ corresponds to an effective attraction that might produce a bound state, the positive sign is characteristic of an effective repulsion and the absence of the bound states[12]. A general feature of the scattering operator is rapid increase of the potential part $t_{pot}(\varepsilon,\theta)$ with the decrease of $\varepsilon$ (or increase of $|\varepsilon|$, since $\varepsilon < 0$) which is due to the enhancement of the exchange interaction of the tunneling electron with the electrons of the scattering center when the energy $\varepsilon$ approaches the energies of the electrons residing on the center[9].

The formulation of the quantum mechanics based on the formalism of scattering operators allows an easy determination of the bound states of the extra electron: the poles of the scattering operator are the energies of the bound states[12]. For a single center the pole term in (3) indicates the existence of a bound state at the isolated center. In the case of a molecular wire that consists of an ensemble of the scattering centers, the bound states of the system can be found as the poles of the total scattering operator $T$. The total scattering operator is easily found as a solution of the system of linear equations:

$$T = \sum_{n=1}^{N} T_n; \quad T_n = t_n + \sum_{k \neq n} t_n G_0(|\mathbf{R}_\mathbf{n}-\mathbf{R}_\mathbf{k}|;\varepsilon) T_k(\varepsilon), \quad n=1,\ldots,N$$
(6)

The roots of the determinant of the system (6) are the poles of $T$. Therefore, the energies of the bound states are determined from the following secular equation:

$$\det|\delta_{ik} - t_i(\varepsilon,\theta)G_0(|\mathbf{R}_\mathbf{i}-\mathbf{R}_\mathbf{k}|;\varepsilon)| = 0, \quad i,k=1,\ldots,N, \quad (7)$$

and they form the band of energy levels. The angles of scattering, $\theta$, in secular equation (7) are chosen according to the paths of the scattering and in the case of a one-dimensional molecular wire they take the values of 0 or $\pi$.

The wave function $\psi_s$ of the tunneling electron corresponding to the $s$-th root of secular equation (7), $\varepsilon_s$, is written as

$$\psi_s(\mathbf{r}) = \sum_l \tilde{T}_l(s)\tilde{\varphi}(\mathbf{r}-\mathbf{R}_l,\varepsilon_s) \quad (8)$$

where $\tilde{\varphi}(\mathbf{r}-\mathbf{R}_l,\varepsilon_s)$ represents additional contributions to the exponential tail of the electron wave function of the tunneling electron due to its interaction with the scattering center $l$ at the position $\mathbf{R}_l$. Their role in forming electron wave functions is similar to the role the atomic wave functions play in building an independent electron, tight-binding wave function (1). The components of the $s$-th eigenvector $\tilde{T}_l(s)$ are obtained as a normalized solution of the homogeneous system (6) and they are similar to the Bloch wave phase factors $\sin(kR_l)$ in the tight-binding Bloch wave function (1).

In accordance with the Bloch theorem, spatial behavior of $\tilde{T}_l(s)$, i.e. the dependence of $\tilde{T}_l(s)$ on the position of the centers $R_l$, is obtained as

$$\tilde{T}_l(s) = A_s \sin(k_s R_l),$$

where the quasi wave-vector $k$ is equal to the number of nodes of the envelope function $\tilde{T}_l(s)$. By examining the spatial behavior, the number of nodes is counted and associated with the appropriate wave vector $k$. In particular, for the case of negative scattering operator $t < 0$ we found that the wave vector $k = (\pi/d)s/(N+1)$, and in the opposite case of $t > 0$, $k = (\pi/d)(N+1-s)/(N+1)$. In order to clarify the general features of the spectrum we will ignore the pole term in (3). Then, a whole set of solutions (8) can be grouped into pairs with identical $\tilde{T}_l(s)$ or the same quasi-wave vectors $k$. Since each solution in the pair has a different eigenvalue, the two states are attributed to two different energy bands that constitute the structure of the energy spectrum of the wire.

In order to show the qualitative picture of the bound state's energy spectrum we will consider the case of a sufficiently large inter-center distance $d$ when $t(\varepsilon,\theta)G_0(d;\varepsilon) \ll 1$ (so-called nearest neighbor scattering). Then, the determinant (7) becomes tri-diagonal: $\det|\delta_{ik} - \beta\delta_{ik\pm1}| = 0$, $\beta \equiv t(\varepsilon,\pi)G_0(d;\varepsilon)$, and can be solved analytically to yield an implicit equation of band energies $\varepsilon$ as a function of the wave vector $k$:

$$\beta(\varepsilon,d) \equiv t(\varepsilon,\pi)G_0(d;\varepsilon) = -\frac{t(\varepsilon,\pi)}{2\pi}\frac{\exp(-\sqrt{2|\varepsilon|}d)}{d} = \frac{1}{2\cos(kd)},$$
(9)

At even larger distances such that $t_{pot}(\varepsilon,\theta)G_0(2d;\varepsilon) \ll 1$, only the pole term of the scattering operator is appreciable. This limit corresponds to the one-electron effective potential approximation and the dispersion (9) becomes identical to tight-binding dispersion (2) with the hopping integral $V = (b(\pi)/d) \cdot \exp(-\sqrt{2|\varepsilon_0|}d)$. This nicely demonstrates the capability of the method to develop a more general picture while establishing connections with previous results.

We will illustrate simple physics of the bound state spectrum by considering the energy dependence of the scattering operator similar to that of triplet scattering of the electron on a hydrogen atom[9]: $t(\varepsilon,\pi)/2\pi = 2.0 - 140.0\varepsilon$. In this case the scattering operator $t > 0$ and the wave vector $k = \pi/d \cdot (N+1-s)/(N+1)$.

It is easy to show that eq. (9) for band energies $\varepsilon$ has a solution only for inter-center distances $d < 4.45$. We calculated the energy spectrum for a particular value of $d = 4$ in nearest-neighbor scattering approximation by solving eq. (9) and solving numerically exact secular equation (7), see Fig. 2. For each value of $k$ there are two solutions of secular equations (7) or (9) which result in two energy bands: upper band $\varepsilon_u(k)$ and the lower band $\varepsilon_l(k)$, see Fig. 2. We also plotted solutions for the case of negative scattering operator $t(\varepsilon)/2\pi = -2 + 140\varepsilon$.

In case of infinite number of centers $N \to \infty$, both bands come together at the energy $\varepsilon = \varepsilon_m$ and the wave vector $k_m$ corresponding to the minimum of the secular curve $t(\varepsilon,\pi)G_0(d,\varepsilon)$ (middle expression in (9)). The dispersion relationship in the vicinity of $\varepsilon_m$ is non-analytic (i.e. $d\varepsilon/dk$ is singular at $k = k_m$): $\varepsilon_{u,l} - \varepsilon_m = \pm a\sqrt{k - k_m}$, where $a = 2/\left[\partial^2 \beta(\varepsilon_m,d)/\partial \varepsilon^2\right]$. The gap between upper and lower bands at $k_m$ in Fig. 2 is due to the finite number of centers in the wire. This dispersion relationship is quite different from the analytic, quadratic dependence of energy on $k$ vector that is characteristic of one-electron effective potential model at the top and the bottom of the tight-binding bands. Moreover, the number of states in each band $N_b$ is less than the number of centers $N$: $N_b = N(\pi - k_m d)/\pi < N$ which is also in contrast to the one-electron effective potential picture of the electron spectrum where the number of energy levels is equal to number of the centers.

For both cases of negative and positive scattering operators, the topology of the dispersion curves is very similar. The only differences are that in the case of negative $t$ the secular curve $\beta(\varepsilon) = t(\varepsilon,\pi)G(d,\varepsilon)$ will lie above the energy axis, the wave vectors will be in the interval $0 < k < k_m \leq \pi/2d$, and the number of the states in both bands will be $N_b/N = k_m/\pi$, see Fig. 2. It is worth noting that in both cases of positive and negative scattering operators the number of bound states in the upper $\varepsilon_u(k)$ and lower $\varepsilon_l(k)$ bands will be less than half of the number of the scattering centers.

The complexity of solutions of the secular equation (7) beyond nearest neighbor scattering do not allow an analytical solution but can be solved numerically. We found that the major features of the bound energy bands remain the same. We can explore in detail the specific features of the energy spectra in the general case and contrast the differences for the systems with negative and positive scattering operators $t$. As has already been mentioned above, the secular equation (7) includes scattering operators at both 0 and $\pi$ scattering angles, but for the purpose of comparison we consider the averaged, angle independent expressions $t(\varepsilon) = 2 - 140\varepsilon$ for $t > 0$ and $t(\varepsilon) = -2 + 140\varepsilon$ for $t < 0$ and $d = 4$.

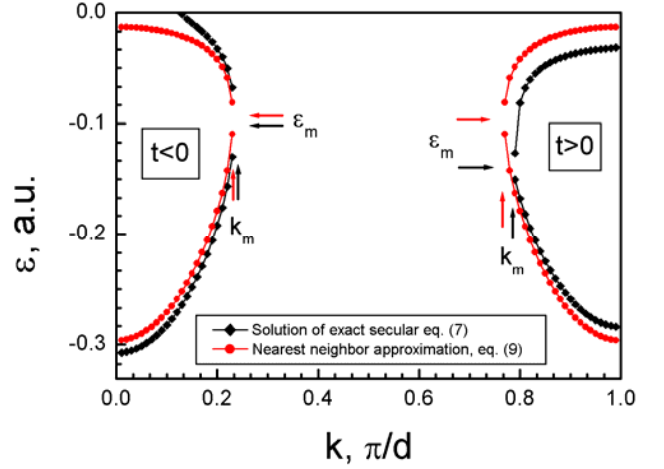

Fig. 2. Band structure of the bound states in molecular wire consisting of $N = 100$ centers separated by distance $d = 4$. Arrows show the centers of the gaps for both exact and nearest neighbor cases.

The numerical solution of the secular equation for both positive and negative $t$ is shown in Fig. 2. We see the reduction of the bandwidth in the case of a positive scattering operator $t > 0$, and an increase of the bandwidth in the case of $t < 0$ as compared to the nearest neighbor scattering approximation, eq. (9). In addition, an asymmetry of the energy spectrum is observed in the case of $t < 0$: the number of states in the upper band is less than that in lower band, see Fig. 2. The upper band has fewer states because the states at the top of the upper band close to zero energy are pushed into the continuum spectrum $\varepsilon > 0$ and become resonant levels there. If the parameters of the interactions in the wire are changed in such a way as to reduce the bandwidth (for example, if $b = 0$ in (3) and $t = 140\varepsilon$), the symmetry of the spectrum is restored because all the states are in the negative energy interval.

As we have already shown, the sign of scattering operator $t$ determines the region in quasi-momentum space $k$ where the energy levels of the bound states exist. Then, substituting parametrization of $k$ at the beginning and the end of Brillouin zone we obtain

$$\tilde{T}_l(s) = \begin{cases} A_s^\nu \sin\left(\dfrac{\pi s R_l}{(N+1)d}\right), & t < 0 \\ A_s^\nu (-1)^l \sin\left(\dfrac{\pi s R_l}{(N+1)d}\right), & t > 0 \end{cases} \quad (10)$$

where $s$ is the index counting energy levels from the bottom of the lowest band ($t<0$) or from the top of the highest band ($t>0$) and the coefficient $A_k^\nu$ depends not only on the wave vector $k$ but also on the band index $\nu$.

In conclusion, we have studied the bound states of the extra electron in a molecular wire, corresponding to the physical situation of the electron tunneling through the wire. Consistent application of the formalism of the scattering operator allows the revealing of unusual properties of the energy spectrum of the extra electron scattering off a linear chain of three-dimensional scattering centers. The developed theory of the bound states is important from a fundamental point of view because it demonstrates substantial deviations from the single-electron effective potential picture of an extra electron in a system with geometrical periodicity. These findings are important for developing physically viable theory of electron tunneling in single molecular systems because the structure of spectrum of electron's bound states and its relative position in respect to the Fermi levels of electrodes determine specific mechanisms of electron transport through a molecular wire.

M.A.K. and I.I.O. thank NAS/NRC for the support within COBASE program and NSF under grant CCF- 0432121. M.A.K. and V.S.P. thank the Russian Foundation for Basic Research for financial support under grant 05-03-32102. I.I.O thanks Carter White for his hospitality during I.I.O.'s stays at Naval Research Laboratory as ONR/NRL Summer Faculty Fellow and Brett Dunlap for discussions.